\documentclass{llncs}
\usepackage{graphicx}
\usepackage{subfigure}
\usepackage{multirow}
\usepackage{multicol}
\usepackage{amsmath}
\usepackage{xspace}
\usepackage{color}
\usepackage{algorithm}
\usepackage{algorithmic}

\begin{document}
%
%\frontmatter          % for the preliminaries
%
\pagestyle{empty}  % switches on printing of running heads
\addtocmark{} % additional mark in the TOC

\def\myalgo{\textsf{EMCD}\xspace}
\def\filter{\textsl{WGP}\xspace}
\newcommand{\M}{\textbf{M}}
\def\G{\mbox{$\mathcal{G}$}} %  
\def\V{\mbox{$\mathcal{V}$}\xspace} % 
\def\E{\mbox{$\mathcal{E}$}} %  
\newcommand{\VM}{$\mathcal{V}_M$}
\newcommand{\EM}{$\mathcal{E}_M$}
\newcommand{\GMDef}{\mbox{$G_M = \langle V_M, E_M, w \rangle$}}
\newcommand{\GM}{\mbox{$G_M$}}
\newcommand{\pvalue}{\mbox{$\gamma_{uv}$}}
\newcommand{\pvalueij}{\mbox{$\gamma_{ij}$}}
\def\lbBaseline{\textsf{CC-EMCD}\xspace}
\def\ubBaseline{\textsf{C-EMCD}\xspace}
\def\greedy{\textsf{M-EMCD}\xspace}
\def\greedystar{\textsf{M-EMCD$^*$}\xspace}

\def\fftwyt{FF-TW-YT}
\newcommand{\algo}[1]{\textsf{#1}}

\newcommand{\argmax}{\operatornamewithlimits{argmax}}

 \mainmatter              % start of the contributions

 \title{Consensus Community Detection in Multilayer Networks using   Parameter-free Graph Pruning}

\titlerunning{}  % abbreviated title (for running head)

 \author{Domenico Mandaglio \and Alessia Amelio \and Andrea Tagarelli}

\authorrunning{D. Mandaglio, A. Amelio, A. Tagarelli} % abbreviated author list (for running head)
%\tocauthor{Domenico Mandaglio, Andrea Tagarelli, Alessia Amelio}
%
 \institute{DIMES - University of Calabria, 87036 Rende (CS), Italy\\
\email{\{d.mandaglio,a.amelio,tagarelli\}@dimes.unical.it}}

\maketitle              % typeset the title of the contribution
\thispagestyle{empty}

\begin{abstract}
The   clustering ensemble paradigm has emerged as an effective tool for community detection in multilayer networks, which allows for producing consensus solutions that are designed to be more robust to the algorithmic selection and configuration bias.  However, one limitation is related to the dependency on a co-association threshold that controls the degree of consensus in the community structure solution.  
 The goal of  this work is to  overcome this limitation   with   a new framework of  ensemble-based multilayer community detection, which  features  parameter-free identification  of consensus communities based on  generative models of  graph pruning that are able  to filter out  noisy co-associations. We also present an enhanced version of the modularity-driven ensemble-based multilayer community detection method, in which community memberships of nodes are reconsidered to optimize the multilayer modularity of the consensus solution. 
  Experimental evidence on real-world networks confirms the beneficial effect of using model-based filtering methods and also shows the superiority of the proposed method on state-of-the-art multilayer community detection. 
\end{abstract}

%\keywords{}

\section{Introduction}\label{sec:intro} 
 Multilayer networks are pervasive in many fields related to network analysis and mining~\cite{Kivela+14,Magnanibook}. 
Particularly, \textit{community detection in multilayer networks} (ML-CD) has attracted lot of attention in the past few years, as witnessed by a relatively large corpus of studies (see, e.g., \cite{KimL15} for a survey).
 
 An effective approach to   ML-CD   corresponds to  \emph{aggregation methods}, whose goal is to  infer a community structure by combining information from   community structures separately obtained on each of the layers~\cite{TangWL09,TangWL12,Tagarelli2017}. % BerlingerioPC13}.
  A special class of such methods resembles theory on \emph{clustering ensemble}~\cite{Strehl2003,GulloTG09}: given a set of clusterings as different  groupings of the input data,  a \textit{consensus} criterion function is optimized to induce a single, meaningful solution that   is representative of the input  clusterings. A key advantage of using a consensus clustering      approach is that the inconvenience of guessing the ``best''   algorithm selection and 
  parametrization   is avoided, and hence   consensus results will be more robust and show higher quality when compared to single-algorithm clustering. 
%   In particular, the \emph{clustering ensemble} methods are based on a set of clustering solutions which define the \emph{ensemble}, detected by different clustering methods or settings on the same set of objects \cite{Strehl2003}, \cite{Ngu2007}, \cite{GulloTG09}. From the \emph{ensemble}, these methods compute a final \emph{consensus} solution. This is a single more stable solution derived from the optimization of an objective function over the information characterizing the clustering solutions in the \emph{ensemble}.  

Despite the well-recognized benefits of using the consensus/ensemble clustering paradigm, 
its exploitation to  ML-CD is, surprisingly, relatively new in the literature~\cite{TangWL12,LancichinettiF12,Tagarelli2017}; 
 actually, to the best of our knowledge, only the most recent of these works goes beyond the use of a clustering ensemble approach as a black-box tool for   ML-CD, by proposing the first well-principled formulation of the \textit{ensemble-based community detection} (EMCD) problem.  Indeed, in~\cite{Tagarelli2017}, aggregation is not limited at node membership level, but it also accounts for intra-community and inter-community connectivity; moreover, the consensus function is optimized via \textit{multilayer modularity} analysis,  instead of being simply  based  on the sharing of a certain minimum percentage of clusters in the ensemble.    %and the consensus solution is discovered from a space of candidates bounded by two community  structures that are representative of the input ensemble.  
 % 
  %Recently, the EMCD method was proposed, extending the concept of \emph{clustering ensemble} for detecting the community structure of multilayer networks \cite{Tagarelli2017}.  Given the single-layer community structure solutions defining the \emph{ensemble}, the method finds the \emph{consensus community structure} optimizing an objective function according to the information from the single-layer community structures. 

 The EMCD method proposed in~\cite{Tagarelli2017} relies on a \textit{co-association-based consensus clustering scheme}, 
 i.e., the consensus clusters are derived from a co-association matrix built to store the fraction of clusterings in which any two nodes are assigned to the same cluster.  
% From this matrix, only meaningful co-association values are retained, by dropping the lowest ones which reflect unlikely consensus memberships, and hence are due to noise. 
 Low values in  this matrix would  reflect unlikely consensus memberships, i.e., noise,  and hence should be removed;  to this purpose, the matrix is subjected to a filtering step based on a user-specified parameter of minimum co-association,  $\theta$. %The main limitation relies on the difficulty of appropriately selecting the $\theta$ parameter. 
% suitably defined over the multiple layers of a network. 
%It takes as input a co-occurrence (or co-association) matrix {\bf M} such that the $(i,j)$-th entry of the matrix stores the number of single-layer community detection solutions in the ensemble in which the $i$-th and $j$-th objects appear in the same cluster, divided by the size of the ensemble. From the co-association matrix {\bf M}, only meaningful co-association values are retained, by dropping the lowest ones which reflect unlikely consensus memberships, and hence are due to noise. Therefore, {\bf M} is subjected to a filtering step based on a user-specified parameter of minimum co-association $\theta \in [0, 1]$. The main limitation relies on the difficulty of appropriately selecting the $\theta$ parameter.
 Unfortunately, setting an appropriate $\theta$ for a given input   network is a challenging task, since too low values will lead to few, large communities, while too high values will lead to many, small communities. 
Moreover, this approach generally  fails to consider      properties related to  node distributions  and linkage  in the network. 

In this work, we aim to overcome the above issue, by proposing a new EMCD framework featuring   a \textit{parameter-free identification of consensus clusters} from which the  \textit{consensus community structure} will be induced. 
 Our   idea is to exploit   a recently developed class of \textit{graph-pruning methods based on generative models}, which are designed to filter out ``noisy'' edges from weighted graphs. 
A key advantage of these pruning models is that they do not require any user-specified parameter, since they enable edge-removal decisions by  computing  a statistical $p$-value for each edge based on a null model defined on the node degree and strength distributions. % --- an edge will be considered as meaningful, hence will be retained in the graph, if its   $p$-value is below the significance level. 
%to each non-zero entry of {\bf M} based on some null models of edge weight distributions and subsequently removing the least significative ones, namely those with highest $p$-value. We consider an edge meaningful if its associated $p$-value is below a certain significance threshold $\alpha$. It is worth noting that a statistical significance threshold cannot be considered as a parameter.
We originally introduce these models to multilayer community detection and propose an adaptation to multilayer networks. 

Another limitation of EMCD is that the community membership of nodes    remains the same through the process of detecting the modularity-driven consensus community structure. In this work, we also address this point,   by defining a three-stage process in the EMCD scheme, which iteratively seeks to improve the  multilayer modularity of the consensus community structure  based on intra-community connectivity refinement, community partitioning, and  relocation of nodes  from a community to a neighboring one. 
%introducing the capability of reconsidering the node membership to communities during the final \emph{consensus community structure}.
%More specifically, we introduce two operations of \emph{splitting}  of communities, and \emph{relocation} of nodes from one community to another community in order to increase the multilayer modularity of the final \emph{consensus community structure}.

%Experimental results have revealed that  \disc{. ......}
Two main findings are drawn from experimental results obtained on real-world multiplex networks:  (i) some of the model-filters are effective in simplifying an input multilayer network to support improved community detection, and (ii) our proposed framework outperforms state-of-the-art multilayer community detection methods according to modularity and silhouette quality criteria. 
%An experiment conducted on both real-world and synthetic network datasets showed interesting results on the use of the proposed method compared with the baseline EMCD, as well as other direct, flattening and aggregation methods for community detection in multilayer networks. 

In the rest of the paper, we provide background on generative-model-based filters and on the existing EMCD method (Section~\ref{sec:background}). Next, we present our proposed framework (Section~\ref{sec:newframework}). Experimental evaluation and results are discussed in Sections~\ref{sec:evaluation} and \ref{sec:results}.  
Section~\ref{sec:conclusion} concludes the paper.

\section{Background}
\label{sec:background}
%\vspace{-2mm}
\subsection{Generative models for graph pruning}

\textit{Pruning} is a graph simplification task aimed at detecting and removing irrelevant or spurious edges in order to unveil some hidden property/structure of the network, such as its organization into communities.  
 A simple technique adopted in weighted graphs consists in removing all   edges having weight below a   pre-determined, global threshold. Besides the difficulty of choosing a proper threshold for the input data, this approach tends to remove all ties that are weak at network level, thus discarding local properties at node level. % and, consequently, destroying the peculiar heterogeneity of edge weights.  
 
A relatively recent corpus of study addresses the task of filtering out ``noisy'' edges from complex networks based on \textit{generative null models}. The general idea is to define a null model based on  node distribution properties, use it to compute a $p$-value for every edge (i.e., to determine the   statistical significance of properties  assigned  to   edges from a given distribution),  and finally filter out all edges having $p$-value above a chosen significance level, i.e.,  keep  all edges that are least likely to have occurred due to random chance. 

Methods following the above general approach have been mainly conceived to deal with weighted networks, so that   the node degree and/or the node strength (i.e., the sum of the weights of all incident edges) are used to generate a model that defines a random set of graphs resembling the observed network.   
One of the earliest methods is the \textit{disparity}  filter~\cite{Serrano+2009}, which evaluates the strength and degree of each node locally.  %The null hypothesis is that the strength of a node is redistributed uniformly at random over the node's incident edges. The disparity filter hence  
 This filter however introduces some bias in that the strength of neighbors of a node are discarded.   
By contrast, a global null model is defined with the \textit{GloSS} filter~\cite{Radicchi+2011},  as it preserves the whole distribution of edge weights. The null model is, in fact,  a graph with the same topological structure of the original network and with edge weights randomly drawn from the empirical weight distribution.  
  % 
%  {\color{blue} Since all edges have the same probability of being assigned a given weight, the statistical test is the same for every edge, and hence this reduces to setting a global threshold (depending on a chosen significance level) for pruning. }
  % 
%
Unlike disparity and GloSS, the null model proposed by Dianati~\cite{Dianati2016}  is  maximum-entropy based and hence unbiased. Upon it, two filters are defined:  the \textit{marginal likelihood filter} (\textit{MLF}), which is a linear-cost method that assigns  a significance score to each edge based on the marginal distribution of edge weights, and  the \textit{global likelihood filter}, which 
%is an ensemble approach that  
accounts for the  correlations among edges. While performing similarly, the latter filter is more costly than MLF; moreover, both  consider   the strength of nodes, but not their degrees.  
Recently, Gemmetto et al.~\cite{Gemmetto+2017} proposed a maximum-entropy filter,   \textit{ECM}, for keeping   only \textit{irreducible edges}, i.e.,    the filtered network will retain only the edges that cannot be inferred from local information.  
% The general goal  is to unveil  the   backbone of non-redundant structures in a complex network, 
 ECM employs a null model based on the canonical maximum-entropy ensemble of weighted networks having the same degree and strength distribution as the real network~\cite{Mastrandrea+2014}. 
%, which allows to overcome redundancy issues that arise in the aforementioned filters. 
 %
 Due to space limits, we report   details of the MLF, GloSS and ECM filters in the {\bf \em Online Appendix} available at  \texttt{http://people.dimes.unical.it/andreatagarelli/emcd/}.

 %==========================================

\subsection{Ensemble-based Multilayer Community Detection}
\label{sec:emcd} 
Let $G_{\mathcal{L}} = (V_{\mathcal{L}}, E_{\mathcal{L}}, \V, \mathcal{L})$ be a \textit{multilayer network} graph, with set of layers $\mathcal{L}= \{L_1, \ldots, L_{\ell}\}$ and set of entities $\V$.  
Each layer corresponds to a given type of entity relation, or edge-label. 
 %Based on the general multilayer network model described in~\cite{Kivela+14},  
For each pair of entity in $\V$ and layer in $\mathcal{L}$,  let $V_{\mathcal{L}} \subseteq \V \times \mathcal{L}$ be the set of entity-layer pairs representing that an entity is located in a layer. 
The set $E_{\mathcal{L}} \subseteq V_{\mathcal{L}} \times V_{\mathcal{L}}$ contains the undirected links between such entity-layer pairs.   
For every layer $L_i \in \mathcal{L}$, $V_i$ and $E_i$ denote the set of nodes and edges, respectively. Also, the   inter-layer edges connect nodes representing the same entity across different layers (monoplex assumption).  
%$V_{L_i} = \{v \in \V \ | \    (v, L_i) \in V_{\mathcal{L}}\} \subseteq \mathcal{V}$ denotes the set of nodes in the graph of $L_i$, and $E_{L_i} \subseteq V_{L_i} \times V_{L_i}$ the set of edges in $L_i$.  
% %
%In order to simplify notations, $V_{L_i}$ and $E_{L_i}$ are also denoted as $V_i$ and $E_i$, respectively. 
% It is worth noting that while entities (i.e., elements of $\V$) do not need to participate in all layers, each entity is present in  at least one layer; 
% %, i.e., $\bigcup_{i \in 1..\ell} V_{L_i}  = \V$.   
% moreover, the only inter-layer edges are considered as ``couplings'' of nodes,  which represent the same entity between different layers.   

%A key concept is the \textit{community structure ensemble} for a given multilayer network. 

%\begin{definition}[Ensemble of community structures]
%Given a multilayer network $G_{\mathcal{L}} = (V_{\mathcal{L}}, E_{\mathcal{L}}, \V, \mathcal{L})$, with $\ell=|\mathcal{L}|$ layers, an \textit{ensemble of layer-specific community structures} for $G_{\mathcal{L}}$ is a set $\mathcal{E}=\{\mathcal{C}_1, \ldots, \mathcal{C}_{\ell}\}$, such that each $\mathcal{C}_h$ (with $h=1..\ell$) is a  a  community structure  of    the   layer graph $G_h$. 
%%\hfill \ \qed   
% \end{definition}

Given a multilayer network $G_{\mathcal{L}}$,   an \textit{ensemble of community structures} for $G_{\mathcal{L}}$ is a set $\mathcal{E}=\{\mathcal{C}_1, \ldots, \mathcal{C}_{\ell}\}$, such that each $\mathcal{C}_h$ (with $h=1..\ell$) is a   community structure  of    the   layer graph $G_h$.  This  ensemble  could be obtained by applying any non-overlapping community detection algorithm to each layer graph. 
%Dependency   between the layer-specific community structures is not assumed.    %Also, the information regarding the particular community detection method or configuration that was used to generate the community structures is not considered. 
%We also assume that   each community structure in the ensemble is  a partitioning of a layer-specific graph, i.e., communities are  disjoint in terms of node memberships.  

Given an ensemble of community structures for a multilayer network, 
the problem of ensemble-based  multilayer community detection (EMCD) is to compute  a \textit{consensus community structure},  as a set of communities that are representative of how nodes were grouped and topologically-linked together over the layer community structures in the ensemble. 
In order to determine the community membership of nodes in the consensus structure, a \textit{co-association}-based scheme is defined over the layers, to detect a clustering solution (i.e., the consensus) that conforms most to the input clusterings. 
 %
%\begin{definition}[Co-association matrix]
%Given a multilayer network $G_{\mathcal{L}} = (V_{\mathcal{L}}, E_{\mathcal{L}}, \V, \mathcal{L})$, and an ensemble of community structures $\mathcal{E}$ for $G_{\mathcal{L}}$,  
%the \textit{co-association matrix} $\mathbf{M}$ is a matrix with size $|\V| \times |\V|$ and such that   the $(i,j)$-th entry stores the number of communities shared by $v_i,v_j \in \V$, subject to the condition  that the two nodes are linked to each other, divided by the number of layers (i.e., the size of the ensemble).
%%\hfill \ \qed   
% \end{definition}
 %
Given   $G_{\mathcal{L}}$, and   $\mathcal{E}$ for $G_{\mathcal{L}}$,  
the \textit{co-association matrix} $\mathbf{M}$ is a matrix with size $|\V| \times |\V|$, whose  $(i,j)$-th entry is defined as $|m_{ij}|/\ell$, where $m_{ij}$ is the set  of communities shared by $v_i,v_j \in \V$, under the constraint that   the two nodes are linked to each other~\cite{Tagarelli2017}.

  EMCD   is modeled in \cite{Tagarelli2017} as an optimization problem in which the \textit{consensus community structure} solution is optimal in terms of \textit{multilayer modularity}, and is to be discovered within a hypothetical space of consensus  community structures that is delimited by  a ``topological-lower-bound'' solution and by a ``topological-upper-bound'' solution, 
 %topologically bounded by a lower-bound and upper-bound solutions.
 for a given \textit{co-association threshold} $\theta$.  
 %The adopted multilayer modularity has been defined in \cite{Tagarelli2017} (see Supplementary material for further details about the multilayer modularity).
 Intuitively, the topological-lower-bound solution   may be poorly descriptive in terms of multilayer edges that characterize the internal connectivity of the communities, whereas the topological-upper-bound solution   may contain   superfluous  multilayer edges connecting different communities. 
  The modularity-optimization-driven  consensus community structure produced by the method in~\cite{Tagarelli2017}, dubbed \greedy, 
  hence produces a  solution that is ensured to have  higher modularity than  both the topologically-bounded solutions.    
%  Details on the multilayer modularity and on the computation of topological-lower-bound consensus are reported in the {\bf \em   
%Online Appendix}.

%\vspace{-2mm}
\section{EMCD and parameter-free graph pruning}
\label{sec:newframework}
%\vspace{-3mm} 
As previously discussed, the \myalgo framework has one model parameter, i.e., the co-association threshold $\theta$, which allows the user to control the degree of consensus required to every pair of nodes in order to appear in the same consensus community.  
Given a selected value for $\theta$ and any two nodes $v_i,v_j$, we say that their community linkage,   expressed by $M(v_i,v_j)$, is considered as meaningful to put the nodes in the same consensus community iff $M(v_i,v_j) \geq \theta$. 

However, choosing a fixed value of $\theta$ equally valid for all pairs  of nodes raises a number of issues. First,  there is an intrinsic difficulty of guessing the ``best'' threshold --- since too low values will lead to few, large communities, while too high values will lead to many, small communities. Second, the approach ignores any property of the input network, and consequently a single-shot choice of $\theta$ may fail to capture the natural structure of communities.  
Of course, to overcome the two issues in practical cases, one could always try different choices of the parameter and finally select the best-performing one (e.g., in terms of modularity, as done in \cite{Tagarelli2017}), but it is clear that the approach does not scale for large networks.  
%In fact, this approach equates significance with the number of shared communities among the layers and it fails to take into account the properties of the network.  Therefore, thresholding \M \space using the parameter $\theta$ systematically discounts nodes which share a low number of communities with other nodes, so it doesn't allow to identify the structures they represent.

It would instead be desirable to evaluate the significance of the co-associations by taking into account the topology of the multilayer network, so that a relatively low value of co-association might be retained as meaningful  provided that it refers to node relations that make sense only for certain layers, while on the contrary, a relatively high value of co-association could be discarded if it corresponds  to the linkage of nodes that have  high degree and co-occur in the same community in many layers --- in which case, the co-association could be considered as superfluous in terms of community structure.    %%ES.  large community in an event-driven Twitter multirelational networks...i due nodi potrebber oessere attori pirnicplai nell'evento.......

In order to fulfill the above requirement, we define a \textit{parameter-free approach to EMCD} that exploits the previously discussed   pruning models. Since such models are only designed to work with (monoplex) weighted graphs, our key idea is to first  infer a \textit{weighted graph representation of the co-association matrix} associated to a multilayer network and its ensemble of community structures, and then apply a pruning model on it to retain only meaningful co-associations.

\begin{definition}[Co-association graph]
\label{def:coassgraph}
Given a multilayer graph $G_{\mathcal{L}}$,  an ensemble $\mathcal{E}$ of community structures defined over it,   and associated  co-association matrix \M,  
 we define the \textit{co-association graph} \GMDef  as an undirected weighted graph such that 
%\VM = \V, $E_M = \lbrace (u,v) \text{ with weight } w_{u,v} \mid M(u,v)\neq 0 \text{ with } w_{u,v}=|m_{uv} | \rbrace$.\\
$V_M = \V, E_M = \{ (v_i,v_j) \ | \ m_{ij} \neq \emptyset, w_{ij} = |m_{ij}| \}$.
\end{definition}

\begin{figure}[t!]
\centering
\begin{tabular}{cccc}
\hspace{-3mm}
\includegraphics[width=3cm,height=2.8cm]{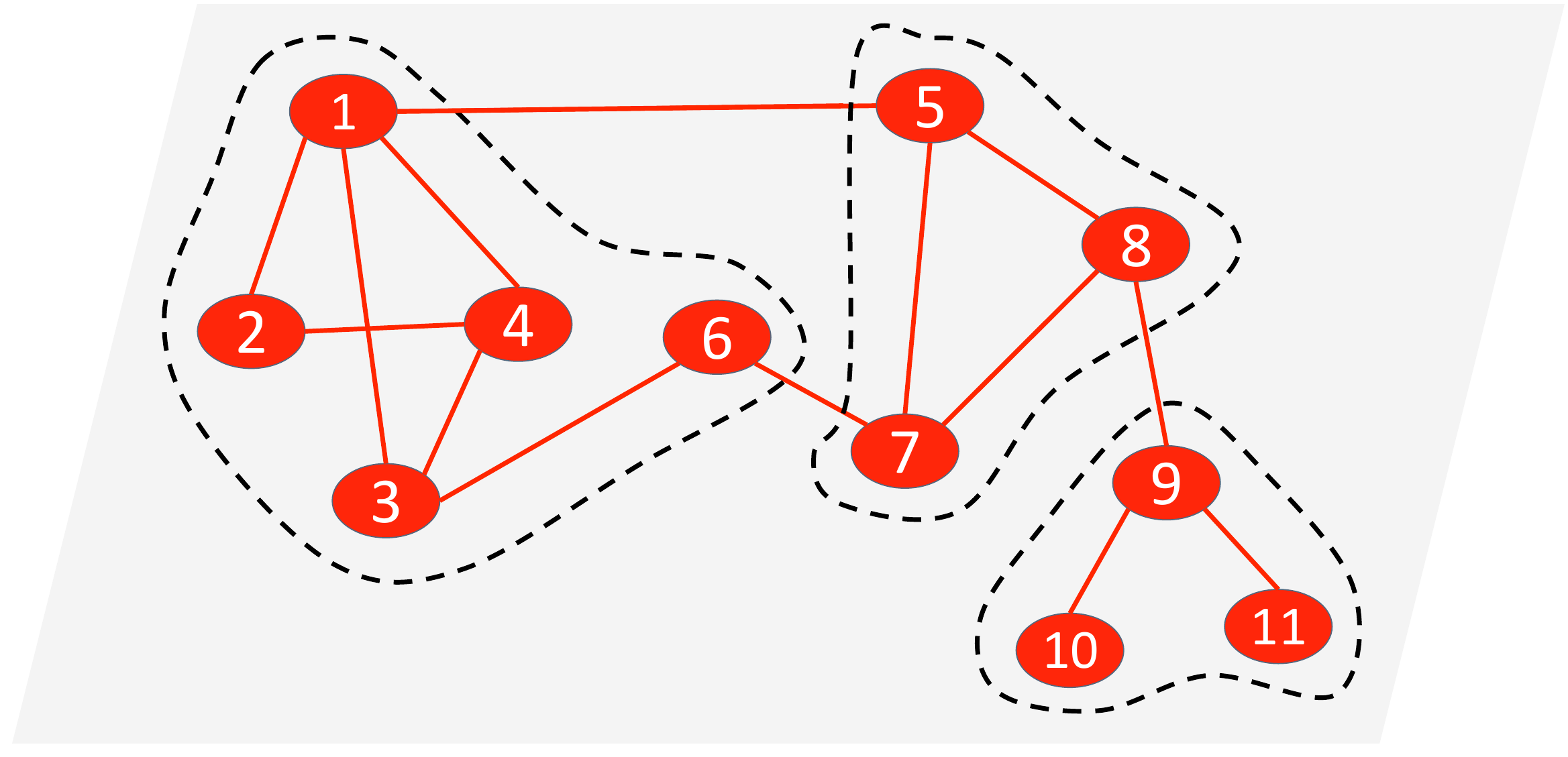} & \hspace{-2mm}
\includegraphics[width=3cm,height=2.8cm]{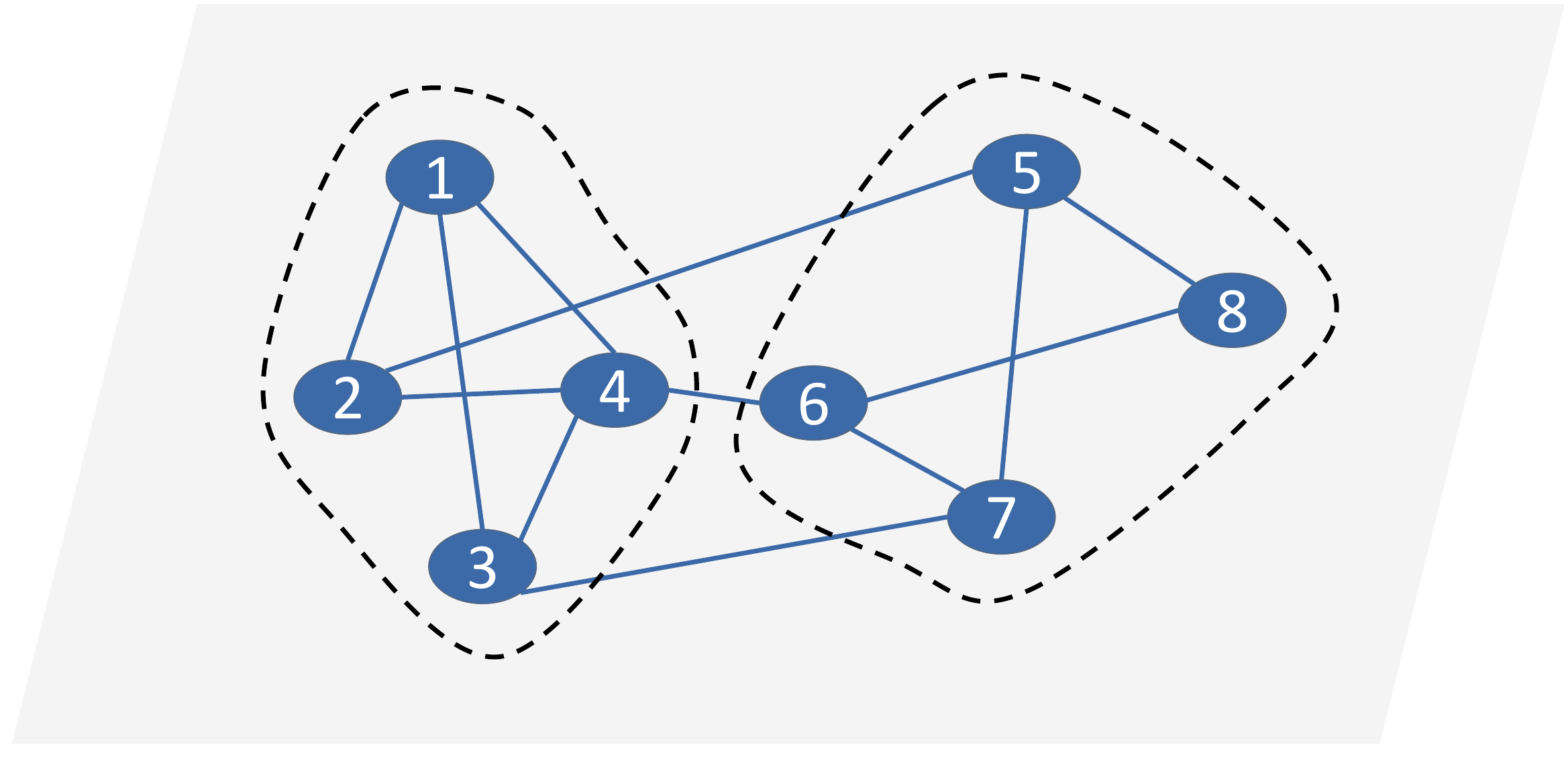} & \hspace{-2mm}
\includegraphics[width=3cm,height=2.8cm]{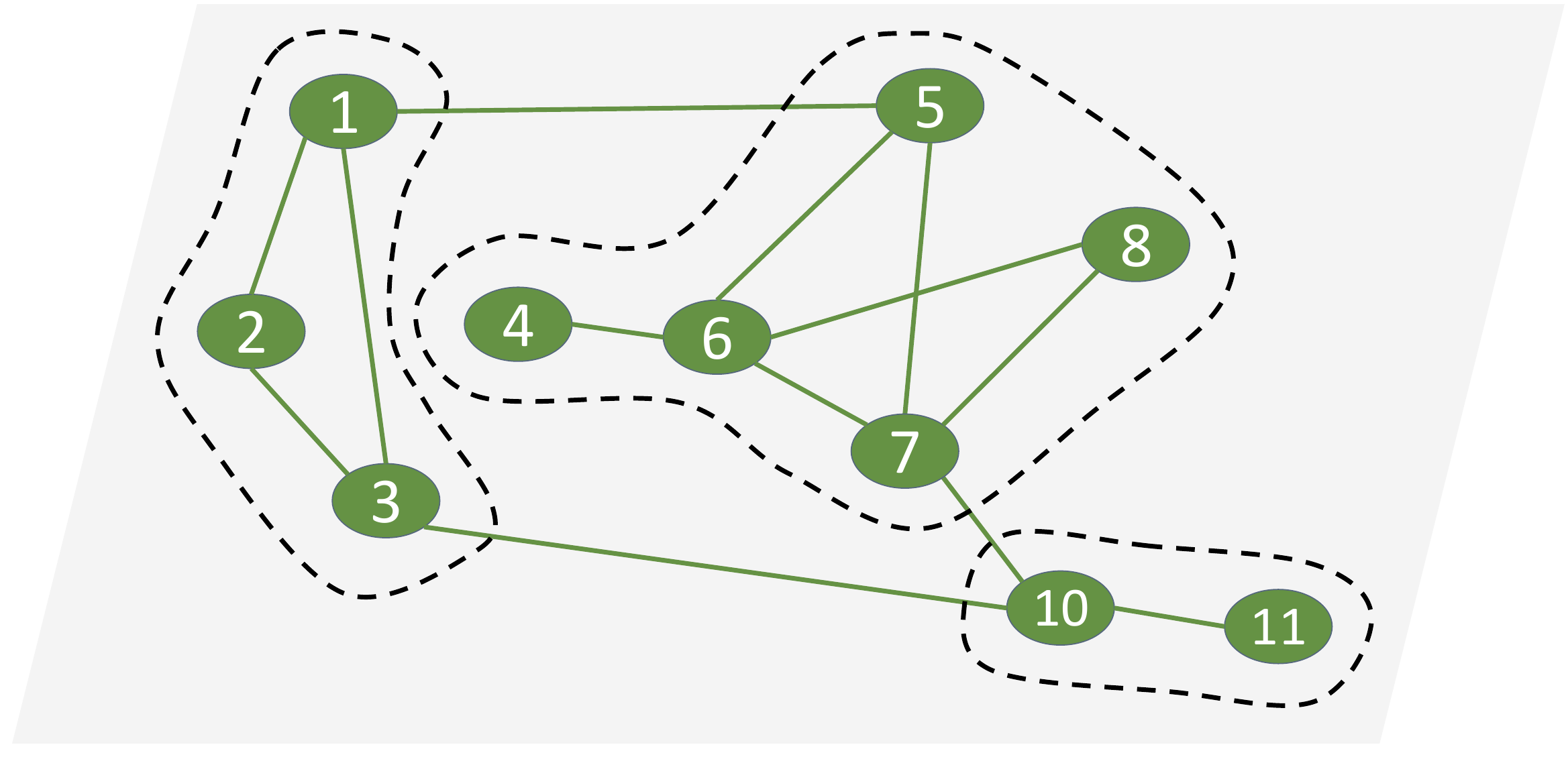} &
%(a) $L_1$ & \hspace{-12mm} (b) $L_2$ & \hspace{-12mm} (c) $L_3$ \\
 \hspace{-2mm} \includegraphics[width=3cm,height=2.8cm]{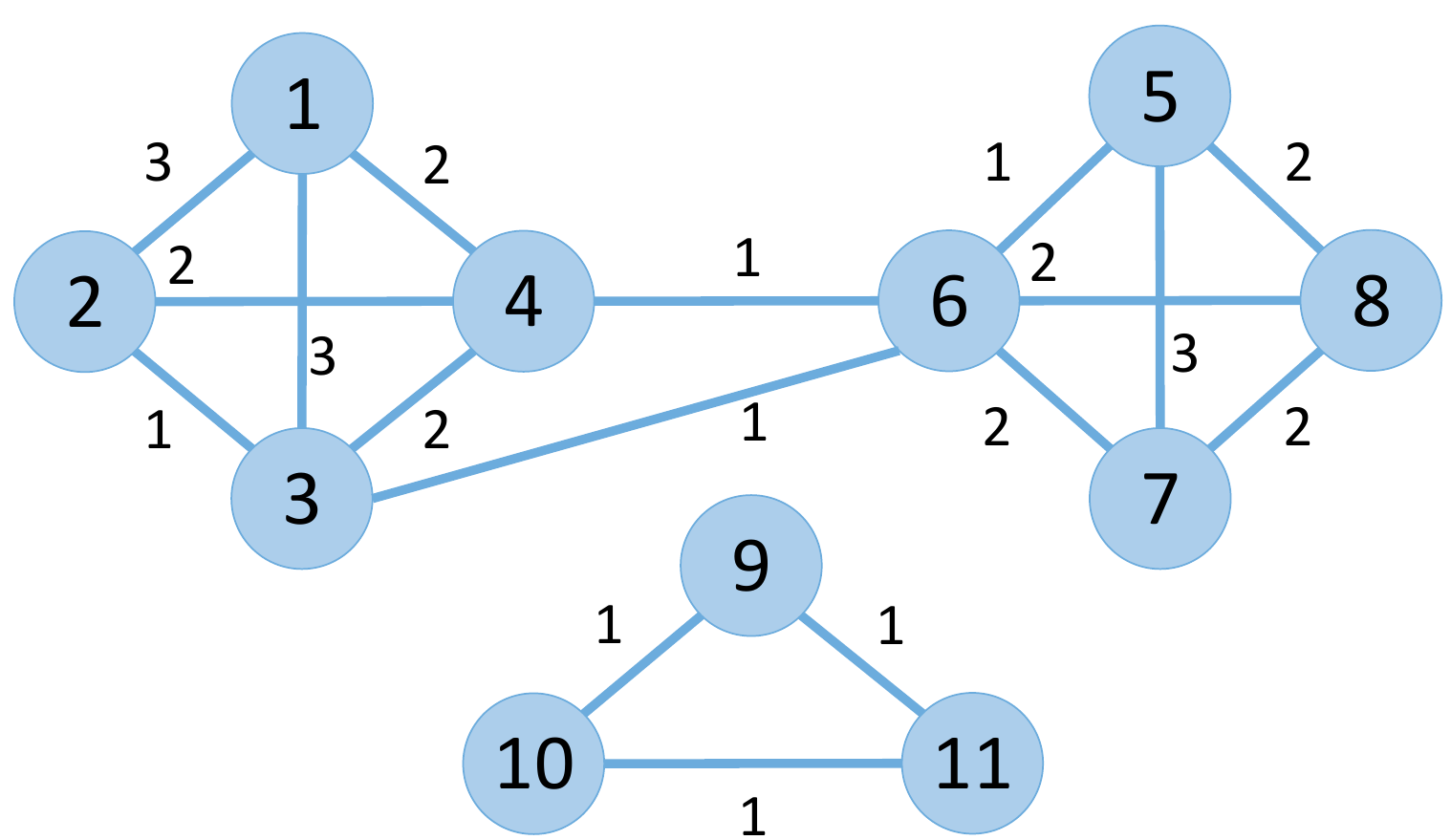}  
% \\ &  \hspace{-12mm} (d) co-association graph  & \  
 \end{tabular}   
%\begin{tabular}{ccc}
%\includegraphics[width=4cm,height=3cm]{img/example/L1_.pdf} & \hspace{-6.8mm}
%\includegraphics[width=4cm,height=3cm]{img/example/L2_.pdf} & \hspace{-6.8mm}
%\includegraphics[width=4cm,height=3cm]{img/example/L3_.pdf} \\
%%(a) $L_1$ & \hspace{-12mm} (b) $L_2$ & \hspace{-12mm} (c) $L_3$ \\
%& \hspace{-6.8mm} \includegraphics[scale=0.35]{img/example/GA.pdf} & \ 
%% \\ &  \hspace{-12mm} (d) co-association graph  & \  
% \end{tabular}   
 \caption{Community structures (denoted by dotted curves) on a 3-layer network, and corresponding co-association graph.}
\label{ex:pruning_es}
\end{figure}

Below is an   example of how the pruning of the co-association graph    based on a user-specified  threshold   could  lead to poorly meaningful consensus communities.

\begin{example}
Consider the 3-layer network and associated co-association graph in Figure~\ref{ex:pruning_es}. 
% shows an example of  how the pruning of the co-association graph matrix  based on a user-specified, global threshold ($\theta$) can lead to poorly meaningful consensus communities.  
Focusing on the community membership of nodes, consider the following settings of  a cutting threshold  $\theta$.  
For  any $\theta \leq 1/3$, all edges will be kept (as the minimum valid weight is 1)  and hence the co-association graph will be partitioned into the two communities corresponding to its two connected components, i.e., $\{1,.., 8\}$ and $\{9,10,11\}$; 
setting  $1/3 < \theta \leq 2/3$ will lead to  $\{1,.., 4\}$, $\{5,.., 8\}$, and $\{9\}, \{10\}, \{11\}$; finally, for 
$2/3 < \theta \leq 1$, the communities will be $\{1,2,3\}$, $\{5,7\}$ and all the other nodes as singletons. 
 It should be noted that no setting of $\theta$ can enable the identification of the three ``natural'' consensus communities, i.e.,   $\{1,.., 4\}$,   $\{5,.., 8 \}$, and $\{9,10,11\}$. 
\end{example}

\begin{definition}[Co-association hypothesis testing]
\label{def:coasstesting}
Given a \textit{co-association graph} \GMDef, let \textsl{WGP} denote a statistical inference method whose generative null model is parametric w.r.t.  node degree and strength distributions in $\GM$.  We define the \textit{co-association hypothesis testing} as a parametric testing based on \textsl{WGP}, whose null hypothesis  for every observed edge is that its  weight has been generated by mere chance, given the empirical strength and degree distributions, and   the associated $p$-value  is  the probability that the null model produces a weight equal to or greater than the observed edge weight.  If the $p$-value is lower than a desired significance level, then the null hypothesis can be rejected, which implies that the co-association of the two observed  nodes is considered as statistically meaningful.   
\end{definition}

 \renewcommand{\algorithmicrequire}{\textbf{Input:}}
\renewcommand{\algorithmicensure}{\textbf{Output:}}
\begin{algorithm}[t!]
	\scriptsize
\caption{Co-association matrix filtering \label{algo:Pruning}}
\begin{algorithmic}[1]
\REQUIRE Multilayer graph $G_{\mathcal{L}} = (V_{\mathcal{L}}, E_{\mathcal{L}}, \V, \mathcal{L})$,  ensemble of community structures  $\mathcal{E}=\{\mathcal{C}_1, \ldots, \mathcal{C}_{\ell}\}$ (with $\ell=|\mathcal{L}|$), generative model for graph pruning \filter.
\ENSURE Filtered co-association matrix $\mathbf{M}$ for $G_{\mathcal{L}}$ and $\mathcal{E}$. \\ 
\STATE Let $\alpha$ be a statistical significance level (i.e., $\alpha=0.05$)
\COMMENT{\textit{Co-association matrix initialization}}
\STATE $\mathbf{M} \leftarrow matrix(|\V|,|\V|)$ 
\FOR{$(i,j) \in \mathbf{M}$}
\STATE $m_{ij} \leftarrow \{ h  \ | \ L_h \in \mathcal{L} \wedge \exists C \in \mathcal{C}_h, \mathcal{C}_h \in \mathcal{E}, \ s.t. \ v_i, v_j  \ in \ C \wedge (v_i,v_j) \in E_h\}$ 
\STATE $M(i,j) \leftarrow  |m_{ij}|/\ell$
\ENDFOR 
%\STATE  $\mathbf{M} \leftarrow$ \textsf{build\_coassociation\_matrix}($G_{\mathcal{L}}, \mathcal{E}$)
% 
\STATE  $\GMDef  \leftarrow$ \textsf{build\_coassociation\_graph}($G_{\mathcal{L}}, \mathbf{M}$)  \hfill 
\COMMENT{\textit{Using Def. \ref{def:coassgraph}}}
\STATE $(e, \pvalueij)_{e=(v_i,v_j) \in E_M}   \leftarrow  $ \textsf{compute\_pValues}($\GM, \filter$)  \hfill 
%\COMMENT{\textit{Using Eq. \ref{eq:pvalueDianati}, \ref{eq:pvalueGloss}, or  \ref{eq:pvalueECM}}}
\COMMENT{\textit{Using Def. \ref{def:coasstesting}}}%
\STATE \textbf{for} $(v_i,v_j) \in E_M$ \textbf{do} 
%\FOR{$(v_i,v_j) \in E_M$}
\STATE \quad \textbf{if}  $\pvalueij \geq \alpha$ \textbf{then} 
$M(i,j) \leftarrow 0$  \hfill 
\COMMENT{\textit{Null hypothesis cannot be rejected}}
%\IF{$\pvalueij \geq \alpha$}  
%\STATE $M(i,j) \leftarrow 0$  \hfill \COMMENT{\textit{Null hypothesis cannot be rejected}}
%\ENDIF
%\ENDFOR
\RETURN{$\mathbf{M}$} 
\end{algorithmic}
\end{algorithm}

Algorithm~\ref{algo:Pruning} shows the general scheme of creation of the co-association matrix, for a given multilayer network and associated ensemble of community structures, and its filtering   based on the co-association hypothesis testing. % defined for a generative model of graph pruning.  

\subsubsection{Enhanced \greedy (\greedystar).\ }
\label{sec:emcd-star} 
We propose an enhanced version of \greedy that has two main advantages w.r.t. the early \greedy method in~\cite{Tagarelli2017}: 
(1) it incorporates   parameter-free pruning of the co-association matrix described in Algorithm \ref{algo:Pruning}, 
and (2)  it fixes the inability of the early \greedy in reconsidering the community memberships of nodes during the consensus optimization.

\begin{algorithm}[t!]
	\scriptsize
	%\footnotesize
\caption{Enhanced Modularity-driven Ensemble-based Multilayer Community Detection (\greedystar)  \label{algo:M*-EMCD}}
\begin{algorithmic}[1]
\REQUIRE Multilayer graph $G_{\mathcal{L}} = (V_{\mathcal{L}}, E_{\mathcal{L}}, \V, \mathcal{L})$,  ensemble of community structures  $\mathcal{E}=\{\mathcal{C}_1, \ldots, \mathcal{C}_{\ell}\}$ (with $\ell=|\mathcal{L}|$), generative model for graph pruning \filter.
\ENSURE Consensus community structure $\mathcal{C}^*$ for $G_{\mathcal{L}}$.  
\STATE $\mathbf{M} \leftarrow \textsf{co-associationMatrixFiltering($G_{\mathcal{L}}, \mathcal{E}$, \filter)}$ \hfill \COMMENT{\textit{Algorithm \ref{algo:Pruning}}} 
\STATE $\mathcal{C}_{lb} \leftarrow \lbBaseline (G_{\mathcal{L}},\mathbf{M})$ \hfill \COMMENT{\textit{Compute topological-lower-bound consensus community structure}} 
%\STATE $\mathcal{C}_{lb} \leftarrow \textsf{lowerBoundWithPruning}(G_{\mathcal{L}}, \mathcal{E}, WGP)$ \disc{Dare altro nome o usare altro alg + lower bound standard}
\STATE $\mathcal{C}^*  \leftarrow  \mathcal{C}_{lb}$
\REPEAT
\FOR{$L_i \in \mathcal{L}$}
\STATE $Q \leftarrow Q(\mathcal{C}^*)$ \\
\COMMENT{\textit{Refine intra-community connectivity of $C_j$}}
\STATE \textbf{for} $C_j \in \mathcal{C}^*$ \textbf{do} 
%\FOR{$C_j \in \mathcal{C}^*$} 
\STATE \quad $\langle C_j', Q_j' \rangle \leftarrow \textsf{update\_community}(\mathcal{C}^*, C_j, L_i)$   
%\ENDFOR
\STATE $j^* \leftarrow \argmax Q_j'$
\STATE \textbf{if} $Q'_{j^*} > Q$  \textbf{then} 
%\IF{$Q'_{j^*} > Q$}
 $Q \leftarrow Q'_{j^*}$,\  $\mathcal{C}^*   \leftarrow \mathcal{C}^* \setminus C_j \cup C'_{j^*}$
%\ENDIF 
\\
\COMMENT{\textit{Refine inter-community connectivity between $C_{j^*}$ and each of its neighbors}}
%\FOR{$C_h \in N(C_{j^*})$}
\STATE \textbf{for} $C_h \in N(C_{j^*})$ \textbf{do} 
\STATE  \quad $\langle \mathcal{C}_h^{IC}, Q_h^{IC}\rangle \leftarrow   \textsf{update\_community\_structure}(\mathcal{C}^*, C_{j^*}, C_h, L_i)$ 
\STATE  \quad  $\langle \mathcal{C}_h^R, Q_h^R\rangle \leftarrow   \textsf{relocate\_nodes}(\mathcal{C}^*, C_{j^*}, C_h)$ 
\STATE  \quad  $\langle \mathcal{C}_h, Q_h \rangle \leftarrow \argmax \lbrace Q_h^{IC}, Q_h^R \rbrace$
%\ENDFOR
\STATE     $h^* \leftarrow \argmax Q_h$
\STATE \textbf{if} $Q_{h^*} > Q$  \textbf{then} 
%\IF{$Q_{h^*} > Q$}
\STATE \quad $Q \leftarrow Q_{h^*},  \  \mathcal{C}^*  \leftarrow \mathcal{C}_{h^*}$
%\IF{$ \mathcal{C}_{h^*}$ has been obtained through relocation}
\STATE \quad \textbf{if} $Q_{h^*} =  Q_{h^*}^R$  \textbf{then} 
%\IF{$Q_{h^*} =  Q_{h^*}^R$}     %\COMMENT{\textit{Relocation case}}  
 $\langle \mathcal{C}_h, Q_h \rangle \leftarrow   \textsf{update\_community\_structure}(\mathcal{C}^*, C_{j^*}, C_{h^*}, L_i)$
%\ELSE 
\STATE \quad \textbf{else} $\langle \mathcal{C}_h, Q_h \rangle \leftarrow \textsf{relocate\_nodes}(\mathcal{C}^*, C_{j^*}, C_{h^*})$
%\ENDIF
%\IF{$Q_h > Q$}
\STATE \quad \textbf{if} $Q_h > Q$ \textbf{then} 
  $Q \leftarrow Q_h, \text{ } \mathcal{C}^* \leftarrow \mathcal{C}_h$
%\ENDIF
%\ENDIF 
\\
\COMMENT{\textit{Evaluate  partitioning of $C_{j^*}$ into smaller communities}}
\STATE $\langle C_s', Q_s' \rangle \leftarrow \textsf{partition\_community}(\mathcal{C}^*, C_{j^*})$  
\STATE  \textbf{if} $Q_s'>Q $ \textbf{then} 
%\IF{$Q_s'>Q $}
  $Q \leftarrow Q'_{s}$, \ $\mathcal{C}^*   \leftarrow \mathcal{C}^* \setminus C_{j^*} \cup C_s'$ 
%\ENDIF
\ENDFOR
\UNTIL{$Q(\mathcal{C}^*)$ cannot be further maximized}
\RETURN{$\mathcal{C}^*$}
\end{algorithmic}
\end{algorithm}

Algorithm \ref{algo:M*-EMCD} shows the pseudo-code of our proposed enhanced \greedy, dubbed \greedystar. 
 Initially, the filtered co-association matrix computed by a selected model-filter \filter is provided as input to \lbBaseline, which computes the initial (i.e., lower-bound)  consensus community structure (Line 2)~\cite{Tagarelli2017}.   
 % (see the {\bf \em Online Appendix} available at  \texttt{http://people.dimes.unical.it/andreatagarelli/emcd/}. ).     
 % 
 This is  iteratively improved   in a three-stage modularity-optimization process: (i)  refinement of connectivity internal to a selected community, (ii) refinement of connectivity between the community and its neighbors also involving relocation of nodes, and (iii) partitioning of the community.
 % in  within-community and inter-community connectivity (as in \greedy ), involving  relocating nodes in its community's neighborhood and by splitting communities in subcommunities. This four-stage optimization is performed iteratively, by analyzing one layer at time, until no further improvement can be done:
 
The within-community connectivity refinement step (Lines 7-10) consists in seeking in the current solution $\mathcal{C}^*$ the community $C_{j^*}$ whose internal connectivity modification leads to the best modularity gain. The internal refinement of a community $C_j$, applied to the layer $L_i$, is performed by   function \textsf{update\_community} (Line 8) which tries to add as many edges of type $L_i$ as possible between nodes belonging to $C_j$, i.e., the set of edges in $E_i$ whose end-nodes  are both in $C_j$ and are not present in the current solution $\mathcal{C}^*$. The function then returns the modified $C_j$ and the updated modularity.

Once identified the community $C_{j^*}$ at the previous step, the algorithm tries to relocate nodes from $C_{j^*}$ to its neighbor communities $N(C_{j^*})$ and/or to refine its external connectivity with them (Lines 11-20). The inter-community connectivity refinement is carried out by  function \textsf{update\_community\_structure} (Line 12) which, for any layer $L_i$ and neighbor communities $C_j$,$C_h$, evaluates the resulting modularity of adding and/or removing edges of type $L_i$ in the  current   consensus $\mathcal{C}^*$ between $C_j$,$C_h$, compatibly with the set of edges of $L_i$ in the original graph.  
%, i.e., the function never tries to add edges that weren't present in the original layer graph.  
The relocation of one node at a time  from $C_{j^*}$ to a neighbor community $C_h$ is evaluated by \textsf{relocate\_nodes} (Line 13)  until there is  no further  improvement in modularity. The ordering of node examination  is determined by a priority queue that gives more importance to nodes having  more edges (of any type) towards $C_h$ than edges linking them to nodes in their current   community in $\mathcal{C}^*$.

The step of partitioning of $C_{j^*}$ into smaller  communities is carried out by   function \textsf{partition\_community} (Line 21). While this can in principle refer to the use of any (multilayer) modularity-optimization-based community detection method, we choose here to focus on the membership of nodes, and hence to devise this step in the simplified scenario of flattened  representation of the consensus community $C_{j^*}$, i.e., a weighted monoplex graph with all and only the nodes belonging to $C_{j^*}$ and weights expressing the number of layers on which two nodes are linked in $\mathcal{C}^*$. Upon this representation, we   apply a graph partitioning method based on modularity optimization (cf. Sect.~\ref{sec:evaluation}) and finally maintain the resulting  partitioning  only if it led to an improvement in modularity.

\section{Evaluation methodology}
\label{sec:evaluation}

\textbf{Datasets.\ }
%We used six real-world multiplex networks for our evaluation, which are   among the most frequently used in recent, relevant studies in multiplex/multi\-layer community detection. 
 Table~\ref{datasets} summarizes main characteristics of the evaluation networks, which were chosen among the most frequently data used in recent, relevant studies on multiplex/multi\-layer community detection.   
 %In the  {\bf \em Online Appendix}, we provide a concise description and additional numerical details on the evaluation networks.  

\begin{table}[t!]
\caption{Main features of real-world  multiplex network datasets used in our evaluation.}
\centering 
\scriptsize
\scalebox{0.85}{
\begin{tabular}{cc}
\begin{tabular}{|l||c|c|c|}
\hline
&\#entities &\#edges &\#layers  \\
& $(|\mathcal{V}|)$ & & $(\ell)$  \\
\hline \hline
AUCS \cite{KimL15} & 61 & 620 & 5   \\
\hline
 EU-Air  \cite{KimL15} & 417 & 3\,588 & 37  \\
\hline
 FAO-Trade    \cite{Domenico2015} &214&318\,346&364 \\
\hline
\end{tabular}
 & \qquad
 \begin{tabular}{|l||c|c|c|}
\hline
&\#entities &\#edges &\#layers  \\
& $(|\mathcal{V}|)$ & & $(\ell)$  \\
\hline \hline
 FF-TW-YT  \cite{Magnanibook} & 6\,407 & 74\,836 & 3   \\
\hline
 London \cite{ZhangWLY16} & 369 & 441 & 3   \\
\hline
 VC-Graders \cite{ZhangWLY16} & 29 & 518 & 3  \\
\hline
\end{tabular}
\end{tabular}
}
\label{datasets}
\end{table}

\textbf{Competing  methods.\ }  
We selected four of the  most representative methods for multilayer community detection:   \textit{Generalized Louvain} (\algo{GL})~\cite{Mucha10},  \textit{Multiplex Infomap} (\algo{M-Infomap})~\cite{Domenico2014},    \textit{Principal Modularity Maximization}~(\algo{PMM}) \cite{TangWL09}, and 
the consensus clustering approach in~\cite{LancichinettiF12} (hereinafter denoted as \algo{ConClus}). Note that the latter two are aggregation-based methods; in particular,  \algo{ConClus} is a simple  approach for consensus clustering in weighted networks. 
%  A concise description of the  competing methods can be found in the   {\bf \em Online Appendix}.

\textbf{Assessment criteria and   setting.\ } 
For the evaluation of the community structures produced by the various methods, we employed the \textit{multilayer modularity} defined in \cite{Tagarelli2017}, the \textit{multilayer silhouette} defined in \cite{Tagarelli2017}, and \textit{NMI}~\cite{Strehl2003}. 

To generate the ensemble for each evaluation network, following the lead of the study in~\cite{Tagarelli2017}, we used  the serial version of the \algo{Nerstrand} algorithm~\cite{LaSalleK15}, a very effective and efficient method for discovering   non-overlapping communities in (single-layer) weighted graphs via modularity optimization.    
 We also used \algo{Nerstrand} for the community-partitioning step in our \greedystar. 
 
As concerns the competing  methods, we used the default setting for \algo{GL} and \algo{M-Infomap}. We varied the number of communities in \algo{PMM}  from 5 to 100 with increments of 5,  %up to  50 (smaller networks) or 100 (larger networks), 
and finally selected the value corresponding to the highest modularity.  Also, 
 we equipped \algo{ConClus} with \algo{Nerstrand} (for the generation of the clusterings),    set $n_p$ to the number of layers,  and varied $\theta$ in the full range (with step 0.01) to finally select the value that determined the consensus clusters with the  highest average NMI w.r.t. the initial ensemble solutions. 
 %We performed step (1) by applying \algo{Nerstrand} on each layer graph, thus generating an ensemble of community structures. We varied the parameter $\theta$ from 0 to 1, with steps of 0.01 for FAO-Trade, 0.02 for EU-Air and 0.2 for the other networks, but we show only results corresponding to the value of $\theta$ that determined the consensus cluster with highest average NMI w.r.t. the initial ensemble solutions.

  More details about the evaluation networks and the competing methods can be found  in the {\bf \em Online Appendix}   at  \texttt{http://people.dimes.unical.it/} \texttt{andreatagarelli/emcd/}.

\section{Results}
\label{sec:results}

\subsection{Impact of model-filters on \greedystar} 

For every network, we analyzed  size,  modularity and silhouette of  the consensus solution obtained before (i.e., at lower-bound \lbBaseline) and at convergence of the optimization performed by  \greedystar, when using either global threshold $\theta$ pruning or one among    MLF, ECM, and GloSS; in the former case,  the value  of modularity refers to  the consensus solution corresponding to the best-performing  $\theta$ value.  Results are reported in Table~\ref{tab:mod}    and discussed next. 
At the end of this section, we also mention aspects related to  time performance evaluation. 
 
 \vspace{-3mm}
\subsubsection{Size of consensus solutions.\ } 
 MLF and ECM tend to produce similar number of communities. %, also compared to the $\theta$-based approach on AUCS, London and \fftwyt. 
  By contrast, GloSS is  in general much more aggressive than the other models, which causes   proliferation of communities in the co-association graph. Also,    the final solution    by \greedystar can differ in size from the initial consensus by \lbBaseline, due to   the optimization of     modularity.

\begin{table}[t!]
\caption{Size and modularity (upper table) and silhouette (bottom table) of   lower-bound  (\lbBaseline) and \greedystar consensus (in brackets, when applicable, the increments over \greedy),  with or without  model-filters.}
\scriptsize
\centering
\begin{tabular}{c}
\scalebox{0.81}{
\begin{tabular}{|c|cccc|cccc||cccc|}
\hline
&\multicolumn{4}{c|}{\lbBaseline modularity}&\multicolumn{4}{c||}{\greedystar modularity}&\multicolumn{4}{c|}{\greedystar \#communities}\\
\cline{2-13} 
&$\theta$-based&MLF&ECM&GloSS&
$\theta$-based&MLF&ECM&GloSS&
$\theta$-based&MLF&ECM&GloSS\\ 
\hline
AUCS&0.60 &	0.68 &	0.66 &	0.21 &	0.86 \textit{(+0.03)} &	\textbf{0.91} &	\textbf{0.91} &	0.25 
&	14&	13&	18&	52\\
EU-Air&0.73  &	0.60 &	0.60 &	0.07	& \textbf{0.91}  &	\textbf{0.91} &	0.90 &	0.09 
&274&	39&	45&	397 \textit{(-2)}\\
FAO-Trade&0.74&0.59&0.30 &	0.20 &\textbf{1.00} &	\textbf{1.00} &	0.99 \textit{(+0.29)} &	0.99 \textit{(+0.56)}
&41 \textit{(+1)}&	1 \textit{(-2)}&	11 \textit{(+3)}&	40 \textit{(-17)}\\
FF-TW-YT&0.48 &	0.44	&0.44 &	0.05 &0.73 \textit{(+0.12)} &	\textbf{0.94}	& \textbf{0.94}	 &0.05 
&119 \textit{(+33)}&	115&	133&	5134	\\	
London&0.89 &	0.85 &	0.85 &	0.41 & 0.90 &	\textbf{0.97} &	\textbf{0.97} &	0.49 \textit{(+0.06) }
&	45&	46&	46&	340 \textit{(-3)}\\	
VC-Graders&0.22 &	0.33	 &0.27 &-0.01 &\textbf{0.88} \textit{(+0.54)}&	0.44 &	0.43 &	0.03 \textit{(-0.01) }
&3 \textit{(-8)}&	16&	17&	26 \textit{(-1)}\\
\hline	
\end{tabular}
}
\vspace{2mm}\\ 
\scalebox{0.85}{
\begin{tabular}{|c|cccc|cccc|}
\hline
&\multicolumn{4}{c|}{\lbBaseline silhouette}&\multicolumn{4}{c|}{\greedystar silhouette}\\
\cline{2-9} 
&$\theta$-based&MLF&ECM&GloSS&$\theta$-based&MLF&ECM&GloSS\\ \hline
AUCS&0.07&	0.23&	0.28&	0.14&0.37 \textit{(+0.01)}&	0.38	&\textbf{0.40}&	0.15\\	
EU-Air&0.01&	0.16&	0.18&	-0.05&0.09&	0.27&	\textbf{0.30}&	0.04 \textit{(-0.02)}\\	
FAO-Trade&-0.06&	0.01&	0.02&	0.01&0.08 &	1.00 \textit{(+0.91)}&	0.06 \textit{(-0.05)}&	0.06 \textit{(-0.05)}\\
FF-TW-YT&0.00&	0.06&	0.06&	0.03&	0.00 \textit{(-0.04)}	&0.15&	0.12	&0.03\\	
London&0.14	&0.06&	0.06&	0.03&0.18&	\textbf{0.20}&	\textbf{0.20}&	0.12 \textit{(+0.04)}\\	
VC-Graders&0.24&	0.20&	0.21&	0.05&	0.52 \textit{(+0.23)}&	0.24&	0.28&	\textbf{0.83} \textit{(+0.77)}\\	
\hline	
\end{tabular}
}
\end{tabular}
\label{tab:mod}
\end{table}

 \vspace{-3mm}
 \subsubsection{Modularity analysis.\ }   
Looking at the modularity results,   besides the expected improvement by \greedystar over  \lbBaseline in all cases, the following remarks stand out. First, MLF and ECM again   behave similarly in most cases, while GloSS reveals to be much weaker; this is clearly also  dependent on the tendency by GloSS of heavily pruning the co-association graph, as discussed in the previous analysis on the size of consensus solutions.  
  Second, using MLF or ECM  leads to higher modularity w.r.t. the best-performing global threshold, in all networks but VC-Graders. % \disc{motivo?}.  
This would support the beneficial effect deriving from the use of a model-filter for the co-association graph matrix; note however that such results should be taken with a grain of salt, since modularity is computed on differently prunings of the same   network.     
 Also,  FAO-Trade  deserves a special mention, since its much higher multigraph density (13.97) and dimensionality (i.e., number of layers) (cf. Table~\ref{datasets}) also  caused a densely connected co-association graph, with average degree of 74, average path length of 1.67, clustering coefficient of 0.64, and 1 connected component. This makes FAO-Trade a difficult testbed for a community detection task, which explains the outcome reported in Table~\ref{tab:mod}:     
11 consensus communities are produced when using ECM, 41 and 40  with $\theta$-based approach and GloSS, respectively, with most of them singletons and disconnected, and even 1 community for MLF.   

It is worth noting that  most of the performance gains by \greedystar over \greedy are obtained for $\theta$-based pruning, but not for model-filter pruning. This   would suggest the ability of \greedystar  of achieving high quality consensus even when a refined model-filter would not be used. 
%without pruning obtains better performances than \greedy without pruning in terms of modularity, silhouette and redundancy. On the contrary, \greedystar and \greedy with pruning do not show meaningful differences. Hence, the pruning phase in \greedy has the positive effect of obtaining the same performances of \greedystar by avoiding the refinement phase.

\vspace{-3mm}
 \subsubsection{Silhouette and NMI analysis.\ }
In terms of silhouette, the use of model-filter pruning  is beneficial to  both \lbBaseline and \greedystar consensus solutions, where the latter achieve significantly higher silhouette in most cases. 
Among the filters, again  MLF and ECM tend to perform closely ---  with a slight prevalence of ECM --- and better than GloSS (except for  VC-Graders, where the number of communities is close to the number of nodes in the co-association graph). 
% Also, \greedystar obtains comparable results with \greedy. 

%Table \ref{nmi1} shows the
We also measured   the NMI of \greedy and \greedystar model-filter consensus solutions vs. the corresponding solutions obtained by $\theta$-based pruning (results not shown). NMI was found   very high (above 0.8, up to 1.0) in  EU-Air, AUCS, and VC-Graders,  around 0.60-0.70 in FF-TW-YT and London, and around 0.40-0.50 in FAO-Trade. Overall, this indicates that   the model-filter pruning has similar capabilities as the best  $\theta$-based pruning  in terms of community membership, though with the advantage of not requiring parameter selection.

\vspace{-3mm}
\subsubsection{Time performance analysis.\ }
Considering the execution time of model-filter pruning (results not shown),   ECM is  in general  more costly than GloSS, and this in turn more costly than MLF. This gap --- at least one order of magnitude --- of  ECM against  the other two filters can be  explained since its higher requirements due to its capability of preserving both degree and strength distributions. 
Details on this evaluation are reported in  the {\bf \em Online Appendix} available at  \texttt{http://people.dimes.unical.it/andreatagarelli/emcd/}.

%==========================================
 
\subsection{Evaluation with competing methods}
\label{sec:competitors}
 Table \ref{tab:GL-PMM-Infomap} summarizes  the  increments  in terms of size,  modularity,   silhouette  (Table~\ref{tab:mod}), and NMI  % (Table~\ref{nmi1}) 
  of \greedystar solutions  w.r.t.  the corresponding solutions obtained by each of the competitors, by varying model-filters. For the NMI evaluation, we distinguished two cases: the one, valid for \algo{GL}, \algo{PMM}, or \algo{M-Infomap}, whereby the reference community structure is the solution obtained by the method in case of  $\theta$-based pruning,   with $\theta$ selected according to the best-modulari\-ty performance; the other one, valid for \algo{ConClus}, whereby we computed the average   NMI over the layer-specific community structures.   
 
This comparative analysis was focused on the impact of using the various model-filters on the methods' performance. To this end, for every   network and model-filter, 
we first  generated an ensemble of layer-specific community structures via \algo{Nerstrand}, then we  built the co-association graph  %\footnote{To make the comparison with \greedystar fair, we maintained the node linkage constraint in the construction of the co-association matrix.}  
   and applied the filter, finally we removed from the original  multilayer network the   edges pruned by the model-filter, before providing it as input to each of the competing methods. % \algo{GL}, \algo{PMM}, or \algo{M-Infomap}. 

 \begin{table}[t!]
\caption{Increments of number of communities, modularity, silhouette and NMI  of \greedystar solutions, by varying model-filters, w.r.t.  corresponding solutions obtained by \algo{GL}, \algo{PMM},   \algo{M-Infomap}, and \algo{ConClus}. }
\label{tab:GL-PMM-Infomap}
\scriptsize
\centering
\scalebox{0.85}{
\begin{tabular}{c}
\begin{tabular}{|c|ccc|ccc|ccc|ccc|}
\hline
&\multicolumn{12}{c|}{Gains by \greedystar vs. \algo{GL}}     \\
\cline{2-13}
&\multicolumn{3}{c|}{\#communities}&\multicolumn{3}{c|}{Modularity}&\multicolumn{3}{c|}{Silhouette} &  \multicolumn{3}{c|}{NMI w.r.t.}\\
&\multicolumn{3}{c|}{ }&\multicolumn{3}{c|}{ }&\multicolumn{3}{c|}{ } &  \multicolumn{3}{c|}{$\theta$-based pruning}\\
\cline{2-13} 
&MLF&ECM&GloSS&MLF&ECM&GloSS&MLF&ECM&GloSS&MLF&ECM&GloSS\\ \hline
AUCS &+6&+8&+48 &	+0.09	&	+0.08	&	-0.39	  &	+0.11&	+0.10&	-0.01 
 & +0.06	& +0.21	& +0.47 \\
EU-Air &-23&-27&+364 &	+0.12	&	+0.11	&	-0.23	 &	+0.26&	+0.29&	+0.08 &  +0.51	& +0.48	& +0.3\\
FAO-Trade &-5&+4&+30 &	+0.53	&	+0.60	&	+0.70	&	+0.97&	+0.07&	+0.07 & -0.55	 & -0.28	& +0.21\\
FF-TW-YT &+111&+130&+5131 &	+0.29	&	+0.27	&	-0.29	 &	-0.07&	-0.10	&-0.05  &  +0.02	& +0.05 &	+0.4\\
London &+23&+23&+318 &	+0.05	&	+0.05	&	-0.42	 	&+0.08&	+0.08&	-0.30 & -0.14	& -0.13	& -0.06\\
VC-Graders&0&+2&+18&	-0.23	 &	-0.26	 &	-0.40	 &	+0.15&	+0.21&	+0.71 & +0.3	& +0.31 &	+0.08\\
\hline
\end{tabular} 
\\
\begin{tabular}{|c|ccc|ccc|ccc|ccc|}
\hline
&\multicolumn{12}{c|}{Gains by \greedystar vs. \algo{PMM}}     \\
\cline{2-13}
&\multicolumn{3}{c|}{\#communities}&\multicolumn{3}{c|}{Modularity}&\multicolumn{3}{c|}{Silhouette} &  \multicolumn{3}{c|}{NMI w.r.t.}\\
&\multicolumn{3}{c|}{ }&\multicolumn{3}{c|}{ }&\multicolumn{3}{c|}{ } &  \multicolumn{3}{c|}{$\theta$-based pruning}\\
\cline{2-13} 
&MLF&ECM&GloSS&MLF&ECM&GloSS&MLF&ECM&GloSS&MLF&ECM&GloSS\\ \hline
AUCS &-1&+4&+38 &	+0.43	&	+0.29	&	0.00	&	+0.12&	+0.13&	-0.04  & +0.24 &	+0.26 &	+0.18\\
EU-Air &-47&-41&+311 &	+0.66	&	+0.65		&+0.04	 &+0.30&	+0.33&	+0.12 & +0.61	& +0.61 & +0.47\\
FAO-Trade &-39&-29&0 &	+0.91	&	+0.90	&	+0.90	&	+1.02&	+0.06&	+0.07 &   -0.61	& -0.4 & +0.06\\
FF-TW-YT &+104&+122&+5123 &	+0.66	&	 +0.60	&	 -0.03	 &	-0.14&	  -0.15&	-0.12 &  -0.1 &	-0.11	& -0.13\\
London &+1&+1&+295 &	+0.26	&	+0.28	&	0.00	&	 +0.03&	+0.03&	-0.02  &  +0.06	& +0.07 &	+0.16\\
VC-Graders &+1&+2&+11 &	-0.05	 &	-0.01	 &	-0.13 	&	+0.25&	+0.27&	+0.95 & +0.24	& +0.2	& -0.29\\
\hline
\end{tabular}
\\  
\begin{tabular}{|c|ccc|ccc|ccc|ccc|}
\hline
&\multicolumn{12}{c|}{Gains by \greedystar vs. \algo{M-Infomap}}     \\
\cline{2-13}
&\multicolumn{3}{c|}{\#communities}&\multicolumn{3}{c|}{Modularity}&\multicolumn{3}{c|}{Silhouette} &  \multicolumn{3}{c|}{NMI w.r.t.}\\
&\multicolumn{3}{c|}{ }&\multicolumn{3}{c|}{ }&\multicolumn{3}{c|}{ } &  \multicolumn{3}{c|}{$\theta$-based pruning}\\
\cline{2-13} 
&MLF&ECM&GloSS&MLF&ECM&GloSS&MLF&ECM&GloSS&MLF&ECM&GloSS\\ \hline
AUCS &+4&+4&+45 &	+0.18	&	+0.23	&	-0.12	 &	+0.17&	+0.11&	+0.11 & +0.48 &	+0.46	& +0.38\\
EU-Air &-255&-251&+167 &	+0.38	&	+0.37	&	-0.20	 &	+0.35&	+0.37&	+0.18 & +0.74	& +0.74 &	+0.56\\
FAO-Trade &0&+10&+39 &	+1.00	&	0.00	&	+0.99	&	+2.00&	+1.06	&+1.06 & 0	& +0.22	& +0.66\\
FF-TW-YT &+113&+130&+5132 &	+0.20	&	+0.24	&	-0.53	 &	-0.15&	-0.15	&-0.23 & +0.4	& +0.3 &	+0.23\\
London &+37&+38&+338 &	+0.52	&	+0.52	&	+0.05	&	+0.21&	+0.20&	+0.12	& +0.39 &	+0.4 &	+0.84\\
VC-Graders &+15&+16&+25 &	-0.49	 &	-0.50	 &	-0.58	 &	+1.24&	+1.28&	+1.83 & +0.66 &	+0.64	& +0.47\\
\hline
\end{tabular}
\\  
\begin{tabular}{|c|ccc|ccc|ccc|ccc|}
\hline
&\multicolumn{12}{c|}{Gains by \greedystar vs. \algo{ConClus}}     \\
\cline{2-13}
&\multicolumn{3}{c|}{\#communities}&\multicolumn{3}{c|}{Modularity}&\multicolumn{3}{c|}{Silhouette} &  \multicolumn{3}{c|}{avg NMI  of}\\
&\multicolumn{3}{c|}{ }&\multicolumn{3}{c|}{ }&\multicolumn{3}{c|}{ } &  \multicolumn{3}{c|}{ensemble}\\
\cline{2-13} 
&MLF&ECM&GloSS&MLF&ECM&GloSS&MLF&ECM&GloSS&MLF&ECM&GloSS\\ \hline
AUCS &
+5 & +9 & +42 &  +0.33 & +0.38 & -0.26   & +0.13 & +0.17 & -0.11 & -0.03 & +0.00 & +0.03 \\
EU-Air & -25 & -18 & +323   & +0.71 & +0.71 & -0.07 &   +0.23 & +0.27 & +0.06   & -0.05 & -0.04 & +0.20 \\
FAO-Trade & -16 & -11 & +21   & +0.59 & +0.77 & +0.74 &   +0.92 & -0.02 & -0.01 &  -0.55 & -0.27 & +0.01 \\
FF-TW-YT & +17 & +74 & +4885   & +0.48 & +0.47 & -0.33 &   +0.15 & +0.12 & +0.02   & -0.06 & -0.04 & +0.18 \\
London & +16 & +21 & +298   & +0.15 & +0.14 & -0.30   & +0.09 & +0.10 & -0.01   & +0.01 & +0.02 & +0.12 \\
VC-Graders & +10 & +10 & +20 & +0.21 & +0.24 & -0.20 &   +0.09 & +0.11 & +0.68 & +0.02 & -0.04 & -0.14 \\
\hline
\end{tabular}
\end{tabular}
}
\vspace{-3mm}
\end{table}

  One general remark is that   \greedystar equipped with MLF or ECM     outperforms all competing methods in terms of  both modularity and silhouette, and tends to produce more communities, with very few exceptions.   
  Concerning NMI results for the first three methods, again the increments by \greedystar are mostly positive, thus implying that model-filter pruning appears to be more beneficial, w.r.t. a global threshold based pruning approach,  for \greedystar than    \algo{GL}, followed by \algo{PMM} and \algo{M-Infomap}.   Also, it is interesting to observe that, with the exception of FAO-Trade for MLF and ECM,  \greedystar has average NMI of ensemble comparable to or even better  than \algo{ConClus}, whose performance values are optimal in terms of NMI (i.e., the  parameter  threshold  corresponded to the best NMI over each network).

\section{Conclusion} 
\label{sec:conclusion}
 
We proposed   a new framework for   consensus community detection in  multilayer networks. This is  designed to enhance the modularity-optimization process w.r.t. existing EMCD method. Moreover, by exploiting parameter-free generative models for graph pruning, our framework  overcomes the dependency on a user-specified threshold for the global denoising of the co-association   graph.

%=============================
%\bibliographystyle{splncs03}      % basic style, author-year citations
%\bibliography{refs}  

%\vspace{-2mm}

\end{document}